\documentclass{article}
\usepackage{graphicx,amsmath,hyperref,latexsym,float,amssymb}
\pdfoutput=1
\usepackage{siunitx}
\usepackage[square,numbers]{natbib}

\DeclareSIUnit\sq{\ensuremath\Box}
\usepackage[font=scriptsize,labelfont=bf]{caption}
\usepackage{hyperref}
\usepackage{authblk}
\title{ Engineering Surface Oxygen Vacancies in $\mathrm{SrTiO_3}$ to Form  a High Mobility  and Transparent Quasi Two dimensional Electron System}
\author{Shyam Sundar Yadav, Shelender Kumar\footnote{ Equal Contributors.}, Pankaj Kumar$^* $  \\ and Ananth Venkatesan\footnote{email : \href{mailto:v_ananth@rocketmail.com}{v\_ananth@rocketmail.com} }}
\date{}
\affil{Department of Physical Sciences , Indian Institute of Science Education and Research Mohali ,
Knowledge City , Sector 81, SAS Nagar,
Manauli P.O. 140306, India}

\begin{document}

\maketitle
\begin{abstract}{
Quasi-two-dimensional electron systems (q-2DES) are formed in various hetero-structures, including oxide interfaces. Oxygen vacancies (OVs) in oxides like $\mathrm{SrTiO_3}$ are known to produce electronic carriers. A novel way to produce $\mathrm{SrTiO_{3-\delta}}$ on the surface using a low-energy $\mathrm{H_2}$ plasma is shown here. It results in a q-2DES with mobility as high as $\mu \sim 20,000 \; cm^2V^{-1}s^{-1}$, displaying
quantum oscillations in magneto-resistance. We can achieve a sharper or weaker confinement potential by adjusting the process pressure. The system with sharper confinement displays clearer
quantum oscillations and Kondo-like temperature dependence
of resistance. OVs close to the surface behaving like  a correlated Anderson impurity is responsible for the Kondo behaviour. Quantum oscillations are less prominent in the weakly
confined system. A cross-over from weak-localization to anti-localization with temperature is seen, but no Kondo behavior. The process also results in a transparent conductor amenable to lithographic patterning. This conductor's standard figure of merit is comparable to poly-crystalline ITO films in the visible regime and extends with similar performance into the  $\lambda$ $\sim 1.5$ $\mu m$ telecommunication wavelength.
}
\end{abstract}
\section*{Introduction}
Fowler and co-workers \cite{Fowler} reported the first quasi-two-dimensional electron system (q-2DES ) in silicon surfaces with confinement-induced Shubnikov-de-Haas oscillations. Since then, a host of hetero-structures-based on GaAs and other semiconductors, as well as the surface of liquid helium \cite{Ando} have also shown a wide variety of magneto-transport phenomena like quantum hall effects\cite{Klitzing,Tsui} and associated quantum phase transitions \cite{Shahar}, Metal to insulator transitions \cite{MIT} and spin-orbit effects \cite{Dyakonov} to name a few key phenomena. Since the discovery of Graphene \cite{graphene} by exfoliation, a large subset of van-der-Waals hetero-structures \cite{Geim} that form q-2DES has been discovered.
Oxide electronics\cite{Ramirez} is a growing field as the rich orbital structure of these materials adds several functionalities.  Since the discovery of  q-2DES using a possibly polar analog of modulation doping by charge donors \cite{Stormer,Tsui} on interfaces of   $\mathrm{LaAlO_3}$ (LAO) and $\mathrm{SrTiO_3}$ (STO) \cite{Ohtomo2004}, there have been extensive studies of similar hetero-structure based q-2DES \cite{Levy} on STO. With intrinsic properties like quantum para-electricity\cite{quantumpara}, potential devices for spin-charge conversion \cite{ferrospin} have been demonstrated. Also, the heavy effective mass of carriers makes STO a model system for the interplay of $ e-e $ interactions re-normalized by strong dielectric constants as well as making practical devices for transparent electrodes\cite{Correlated}. \\
Apart from hetero-structures, surface doping achieved by ionic gating has also yielded high mobility q-2DES samples \cite{Goldhaber}. Recently  STO epi layers delta doped with Lanthanum has shown quantum Hall effect\cite{Matsubara}. Other in-situ processes  q-2DES for photo-emission studies  producing include intense UV-radiation \cite{mixed}on STO surfaces in Ultra high vacuum (UHV) and adsorbing atomic hydrogen\cite{atomic_H}.

In this work, we have come up with a process that can make delta dope the surface by producing $\mathrm{SrTiO_{3-\delta}}$ on the surface of STO with a low-damage plasma process consistently yielding mobilities as high as
$\mu \sim 20,000 \; cm^2 V^{-1}s^{-1}  $ at low temperatures and displaying clear quantum oscillations of magneto-resistance.

There are no solid-state systems that are free of defects. Defects are sources of scattering of carriers, reducing mobility, but they are not always undesirable.  Other
than phenomena like colour centres defects like vacancies can also  induce the doping of
semiconducting materials\cite{Seitz}. Isolated  nitrogen vacancy centers (NV)  in diamond are suitable for quantum computing protocols\cite{Diamond}. Analogous to NV centers, Oxygen vacancies  OVs play a central role in the electronic properties of STO and a related class of $\mathrm{ABO_3}$ ceramics.

It is well known that the  removal of Oxygen from STO to form $\mathrm{SrTiO_{3-\delta}} $ results in doping. Annealing the bulk wafer in a vacuum \cite{reduction1} produces greyish or black samples with conduction electrons. Argon ions in ion milling systems were known to strip the weakly bound oxygen atoms creating conductive surfaces similar to $\delta$-doping of the surface\cite{dresslhaus,Reagor,Cathode}. So far, only Molecular beam epitaxy or Pulsed Laser Deposition growth devices have shown any quantum oscillations in $\mathrm{SrTiO_3}$ \cite{Levy}. Hetero-structures of $\gamma -\mathrm{Al_2 O_3} $ grown using Ultra-high vacuum pulsed laser deposition techniques have achieved record mobilities above $\geq 100,000 \; cm^2V^{-1}s^{-1} $ \cite{Pryds}. Whether oxygen vacancies are intentionally created or partly unintentional due to device processes, they are the subject of intense study as they are sources of $e-e$ correlations or localized orbital correlations, localized deep gap states, and also a source of weak magnetic phenomena \cite{OV_correlated,spaldin}.

\section*{Device Fabrication}
In this work, we have developed a novel method to produce oxygen vacancies OVs close to the surface with minimal damage to the lattice. Unlike the well-known Argon plasma-based ion milling or reactive ion etching, which is a physical process that produces a dark blackish colour when doped heavily \cite{Bruno,KTO}, our process is a gentler  chemical process involving $\mathrm{H_2}$ plasma. \footnote{IISER Mohali has applied for a patent for this new doping process that produces optically transparent conductors. Provisional patent number:  TEMP/E-1/68575/2021-DEL)  }
This results in $\mathrm{SrTiO_{3-\delta}}$ on the surface, similar to delta doping with an epi-layer with an Oxygen scavenging layer like Aluminium or  Niobium that reduces the oxygen \cite{Rodel,Hwang_Nb}.
Since it is a light gas with an affinity for oxygen, the surface damage is minimal for this process resulting in high mobilities. While mobility of $\mu \sim 20,000 \; cm^2V^{-1}s^{-1} $ at low temperatures may not be record mobility comparable to some hetero-structures based on $\mathrm{Al_2 O_3}$\cite{Pryds} this is still higher by an order of magnitude compared to many standard hetero-structures\cite{Levy}. The simplicity of the process compared to MBE or PLD techniques and the possibility of patterning via lithography makes it a very attractive alternative for both mesoscopic physics and practical devices. Also, the samples are not capped with any dielectric, and the bare surface exposed to air is cooled down in modestly high vacuum with pressures of $P \sim 2\times 10^{-2} \;mbar $ which eventually reaches $P \sim 2\times 10^{-7} \;mbar $ at low temperatures due to cryo-pumping.  An ultra high vacuum system can be expected to out-gas adsorbates from the surface and possibly  yield better mobilities.   A Sentech (SI-500) Reactive Ion etch(RIE) / Inductively coupled plasma (ICP) etch system was used in RIE mode. A $\mathrm {H_2}$ flow rate of $\sim 95\; SCCM $  was maintained.The threshold power to produce noticeable doping in small test pieces was above $15 \;Watts$  of RIE power. This resulted in samples that degraded fast, hence a minimum power of $20\; Watts$ was chosen. Samples doped for around $30\; s$ were stable for several days and remained measurable for several months when kept in a vacuum at cryogenic temperatures.
While some samples were bombarded with the edges covered, a silicon thin-film was used as an outer mesa forming a Hall bar in one sample. Details of the stencil protocols are given in the supporting information (SI) along with data from another sample.

For  quantum transport studies, single-side polished $\mathrm{TiO_2}$ terminated STO procured from Shinkosha Corp was used  with doping times as low as $30\;s$. For our optical studies, we used two-side polished wafers with $\mathrm{TiO_2}$ termination on the device side with a doping time of several minutes. In this recipe, the pressure of the reactive gas is the key parameter that tunes the properties of the  q-2DES formed at the surface.
We present the magneto-transport results from two samples. {\bf Sample-A}: A silicon film defining a Hall bar stencil was bombarded at a pressure of $70\; \mathrm{mTorr}$  and power of $20 \; \mathrm{W}$ for 30 seconds.  In {\bf Sample-B}, the edges were coated with nail varnish before bombardment. A pressure of $10 \; mTorr$ was used with an RF  power of $ 20\; \mathrm{W}$. The DC bias parameters did not change and were around $100 \; \mathrm{V} $ for both these processes. However, we can expect the low-pressure recipe to dope deeper than the high-pressure case where energy per ion is higher and mean-free path of ions are higher.  Both samples used $\mathrm{Ti/Au}$
films as ohmic contacts.
{\bf Sample-C} for optical studies used a repetition of the recipe for sample-A for longer duration. The edges were scratched to avoid non planar doped short circuit paths to the XY plane.
Another sample doped for $30\; s$ where the pressure around $70 \; \mathrm{mTorr} $ fluctuated during processing is discussed in SI. A fourth sample {\bf Sample-D } was patterned using PMMA as a stencil for e-beam lithography to make transparent interconnects.

\section*{Quantum Transport }

Both samples were measured in a cryo-free dilution fridge using standard low-frequency lock-in  measurement techniques in the linear response regime with currents up to $ 0.5-5 \; \mathrm{\mu A} $. Both samples show quantum oscillations in magnetic fields perpendicular to the sample. We obtained an areal  carrier density of $ 6.9 \times 10^{13}\; cm^{-2}$ for  sample A and $ 4.3 \times 10^{14}\; cm^{-2}$  for sample B from Hall measurements. \\
In sample-A, the quantum oscillations at low temperatures are clearly visible in the raw data whereas in sample-B the background polynomial magneto-resistance needs to be subtracted to clearly infer the presence of oscillations.
At these temperatures, sample-A's Hall response was linear, showing it is a single band transport. While most standard $\mathrm{LAO/STO}$ hetero-structures\cite{Joshua}  or liquid ion gated surface states \cite{Goldhaber} show two-band transport at densities well above $ 2 \times 10^{13} \; cm^{-2} $. Superconductivity was not seen in either samples at the lowest temperatures achieved $T\sim 100\; mK $ or $T\sim 50\;mK$ for different cool-downs.

Sample-B shows a higher slope in Hall effect at higher fields. The small deviation at lower field is not described by two bands as seen for LAO/STO\cite{Joshua}. In two-band scenario the slope is stronger at lower fields  due to low density of one sub-band being probed  and weaker in higher fields multiple bands  with  an overall higher density respond. This behaviour is due to dilute magnetic moments\cite{WAL_TI}  possibly due to  the OVs diluted over a large volume  \cite{OV_correlated}. This aspect is also discussed later.

\begin{figure}[H]
\center
\includegraphics[width=\textwidth]{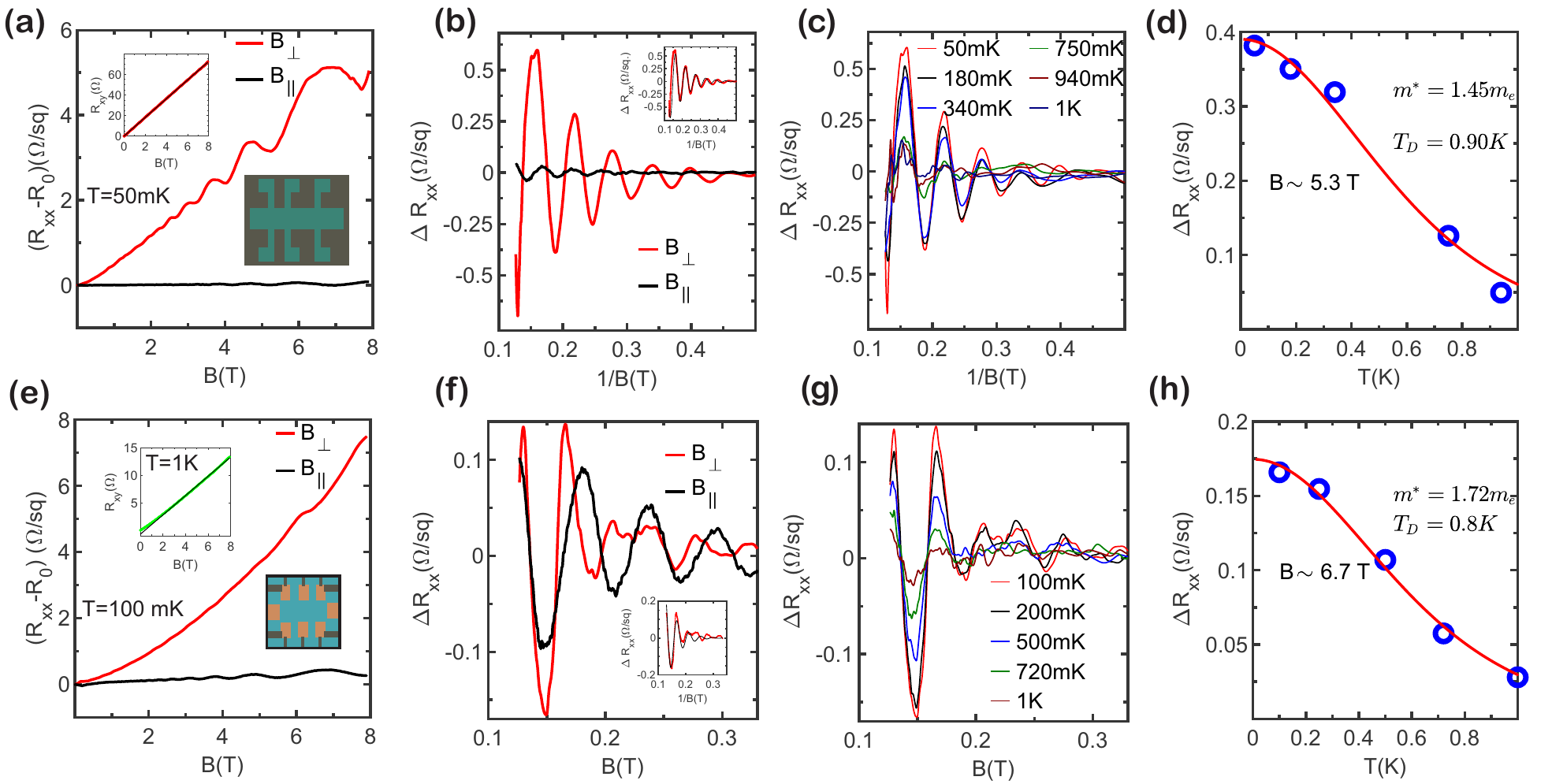}
\caption{{ (a) Strong Shubnikov-deHass(SdH) Oscillations in sample-A doped with a silicon hall bar stencil  at a high pressure of $P \sim 70 \; mTorr$ is visible in raw data $R_{xx}(B)-R_{xx}(0)$ with $R_{xx}(0)\simeq 4.5\; \Omega/_\square$. The low temperature density estimated from the Hall effect is $6.9\times10^{13}\;cm^{-2} $ and mobility $\sim 20015\;cm^2 V^{-1} s^{-1}$. The Hall effect is shown in the inset.
(b) Same data with background subtracted showing a clear Shubnikov-de Hass oscillations. The inset shows a Lifshitz-Kosevich fit to the S-dH oscillations. In a B field parallel to the sample plane, the oscillations are an order of magnitude lower with only few high field peaks visible.
(c) S-dH oscillations with background and Offsets removed for different temperatures (d) A Dingle fit to the temperature dependence of amplitude at 5.3 T. An effective mass of $m^* = (1.45 \;\pm 0.1)m_e$ and a scattering time of $ \tau_s \sim 1.34 \; ps $ is estimated from the fits.
(e) Weak SdH oscillations in sample-B doped at a low pressure of $P \sim 10 \;mTorr$ with the density of the order of $4.3\times10^{14}\;cm^{-2}$ and mobility $\sim 16896\;cm^2 V^{-1} s^{-1}$ estimated at low temperature. The oscillations in raw data are not clear as in sample-A despite of lower zero-field resistance of $R_{xx} \simeq  1\;\Omega/_\square $. The Hall effect is shown in the inset with a higher slope in higher magnetic fields. (f) The $\Delta R_{xx}$ shows the S-dH oscillations at T$\;\sim 100 \; mK$. A higher order(third-order) polynomial is fitted to extract the oscillating part of the magneto-resistance. In a parallel field, the S-dH is slightly phase shifted but with comparable amplitude, indicating a more three-dimensional nature of the electron gas. The LK fit is not as good as the previous case which is expected to have a sharper confining potential. (g) Temperature dependence of the S-dH oscillations (h) Dingle plot for the sample doped at lower pressure at 6.7 T. An effective mass of $m^* =(1.72 \;\pm 0.1)m_e$ and a scattering time of $ \tau_s \sim 1.58 \;ps$ is estimated from the fits.}}
\label{fig1}
\end{figure}

As shown in figure(\ref{fig1}), we see very clear Shubnikov-deHaas (S-dH) type oscillations in all the samples that we studied at low temperatures around $T\sim 100 \;mK$ or slightly lower up-to $50\; mK $. It is well known that in most oxide hetero-structures, unlike systems like silicon \cite{Ando} the oscillations do not go to a zero resistance state as the Landau levels are not fully resolved due to high carrier density and higher effective mass.
The oscillations fade away above $ 1\; K$. As it is common for oxide hetero-structures\cite{Shalom,Caviglia} the density obtained by an analysis of S-dH oscillations is two orders of magnitude lower than the density from Hall effect when we consider spin degeneracy and no valley degeneracy.

Strictly speaking, the Heisenberg uncertainty principle does not allow pure two-dimensional systems. The wave functions of electrons quantized in the z-direction display characteristics of sub-bands \cite{Ando} in the confinement potential. Even quasi-2D materials like graphene have many ripples that are out of plane. Further in oxide hetero-structures, the orbital nature of itinerant electrons stemming from the Ti$\;3d$ orbitals especially $d_{xz},d_{yz} $ shows quasi 3D behaviour even in remotely doped hetero-structures and in UV-excited surface states \cite{mixed}.
In parallel magnetic fields, the samples doped at low pressure, we see S-dH oscillations after removing the background figure(\ref{fig1}f) are as strong as in the perpendicular field. Also, the relative oscillation strength of S-dH in $B_\perp$ is one order of magnitude smaller than the sample doped at high pressures(as seen clearly in figure \ref{fig1}a and \ref{fig1}e ) . This indicates clearly that the low-pressure doped samples are doped deeper as expected with more energy per ion. The absence of oscillations may not always signify ideal 2-D nature; it might also simply imply a poor interface quality. Strangely in superconducting LAO/STO interfaces an analysis of $T_c$ in parallel magnetic field seems to give always a width $\sim 8-10 \; nm$ irrespective of gating voltages \cite{thickness}. Other low density q-2DES show magneto-resistance due to spin polarized states in parallel fields\cite{Okamoto}. These samples have very high density to see any signatures of such phenomena.  In the sample-A doped at high pressure, the oscillations in an in-plane field are shifted weakly due to the orbital character and is an order of magnitude less than the perpendicular field, indicating it is a better quasi-2D system than the previous case.

Temperature dependence of the oscillating part of the resistance are shown in figure(\ref{fig1}.c). Decay of the oscillation amplitude can be used to extract the quantum transport parameters using  Lifshitz-Kosevich(LK) formula\cite{Shalom,shoenberg},
\begin{equation}
\Delta R_{xx}=4R_{c}R_{T} R_{D} sin \left [ 2\pi \left ( \frac{F}{B} -\frac{1}{2} \right )\pm \frac{\pi}{4} \right ]
\end{equation}\label{eq:1}
where $R_{c}$ is the non-oscillating part of the resistance. Temperature-dependent part of the oscillations decays as	$R_{T}=\frac{2\pi^{2}m^{*}k_{B}T}{\hbar eB}/sinh(\frac{2\pi^{2}m^{*}k_{b}T}{\hbar eB})$.
Here $m^{*}$ is the effective mass of the electron, $\hbar$ is the reduced Planck's constant and $e$ be the charge on the electron. The Dingle factor $R_{D}$ is $R_{D}=exp(-\frac{\pi}{\omega_{c}\tau_{D}})$,
where $\omega_{c}=eB/m^*$ is the cyclotron frequency, and $\tau_{D}$ is the dingle scattering time or quantum scattering time and is related to the Dingle temperature, $k_{b}T_{D}=\frac{\hbar}{2\pi\tau_{D}}$. For sample A which have more pronounced S-dH oscillations,
the estimated value of effective mass is found to be $m^*\simeq1.45\;m_{e}$, $m_e$ is the free electron mass and  $T_{D}\simeq0.9\;K$. Effective mass is comparable to the other STO-based quasi two-dimensional electron gas\cite{Caviglia,Hwang_Nb}.  We get  an effective quantum mobility, $\mu_q=\frac{e\tau_D}{m^*}$ of $\simeq\;8 \% $ of the total Hall mobility. In the inset of figure(\ref{fig1}b) and  figure(\ref{fig1}f) a fit to the LK formula are shown. For Low pressure doped sample LK fit is not as good as the high pressure sample  due to very weak oscillation amplitudes.

\section*{Kondo like resistance in high pressure doped samples }
\begin{figure}[H]
\centering
\includegraphics{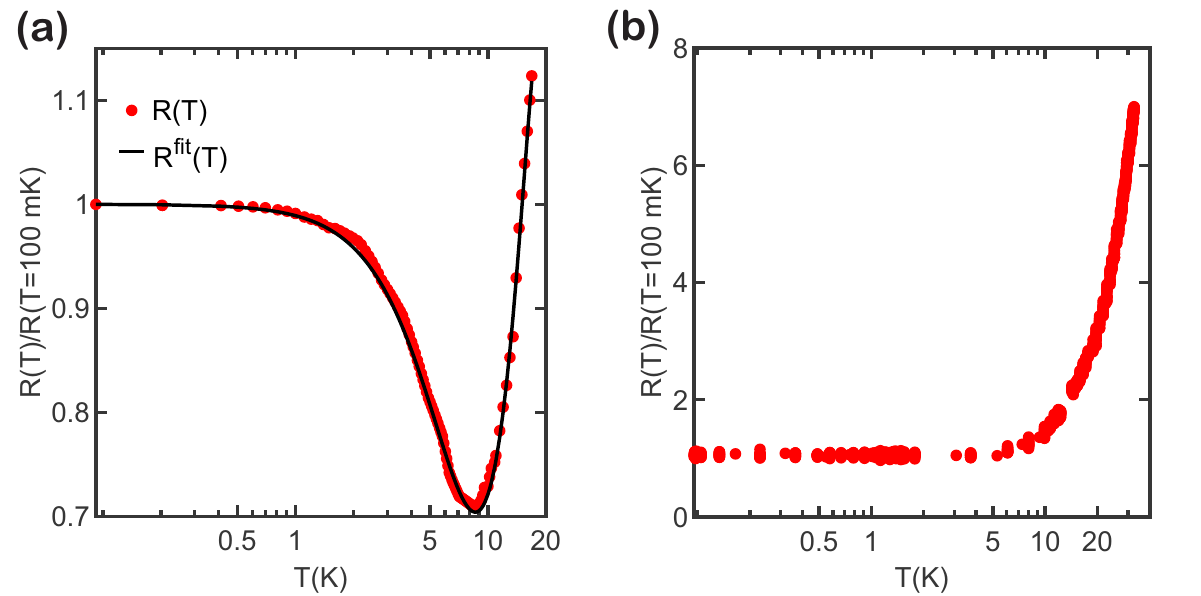}
\caption{ {(a) Resistance vs Temperature  for sample-A showing strong Kondo type upturn behaviour. Black solid line is a fit to the Kondo model(described in SI ). (b) The characteristic  Kondo type upturn in resistance is not seen in the low pressure doped sample (Sample-B).}}
\label{fig2}
\end{figure}
In sample-A doped with high pressure, the  resistance as a function of temperature displays a classic upturn followed by a saturation $ dR/dT \sim 0 $ at low temperatures. Earlier experiments on q-2DES produced by ionic gating modeled this as a Kondo scattering, possibly due to the $\mathrm{Ti^{3+}}$ orbitals\cite{Kondo1}. These experiments showed stronger Kondo-type behaviour in large gate voltages, allowing the electron gas to scan a larger volume of $\mathrm{Ti^{3+}}$ sites. The same phenomena were opposite in LAO/STO hetero-structures\cite{JoshuaKondo} possibly because OVs are closer to the interface. By fitting the data to a standard Kondo-like model, we get a Kondo temperature of $T_K \sim 38\; K $.
Obviously; our high-pressure doping recipe produces a sharper confinement potential at the surface that is more quasi-two-dimensional in nature. The low-pressure recipe, which produces a weaker confinement potential, does not show any Kondo-like upturn behaviour in resistance.
The Kondo-type behaviour is not necessarily universal in  different interfaces. Liquid ion gated samples show a broad Weak anti-localization behaviour in magneto-resistance, indicating magnetic scattering with positive magneto-resistance\cite{Kondo1} along with a non-linear Hall effect indicative of two band scattering \cite{Goldhaber}.
The classic Kondo resistivity minimum is is seen in metals like gold or copper with dilute magnetic impurities. Even thogh the system may not show a strong paramagnetic behaviour it is the dilute local magnetic moments that matter.

Few systems have correlated a Kondo like resistivity upturn to lattice mismatch during growth of hetero-structures\cite{Kondo_lattice}
\begin{figure}[t]
\includegraphics[width=\textwidth]{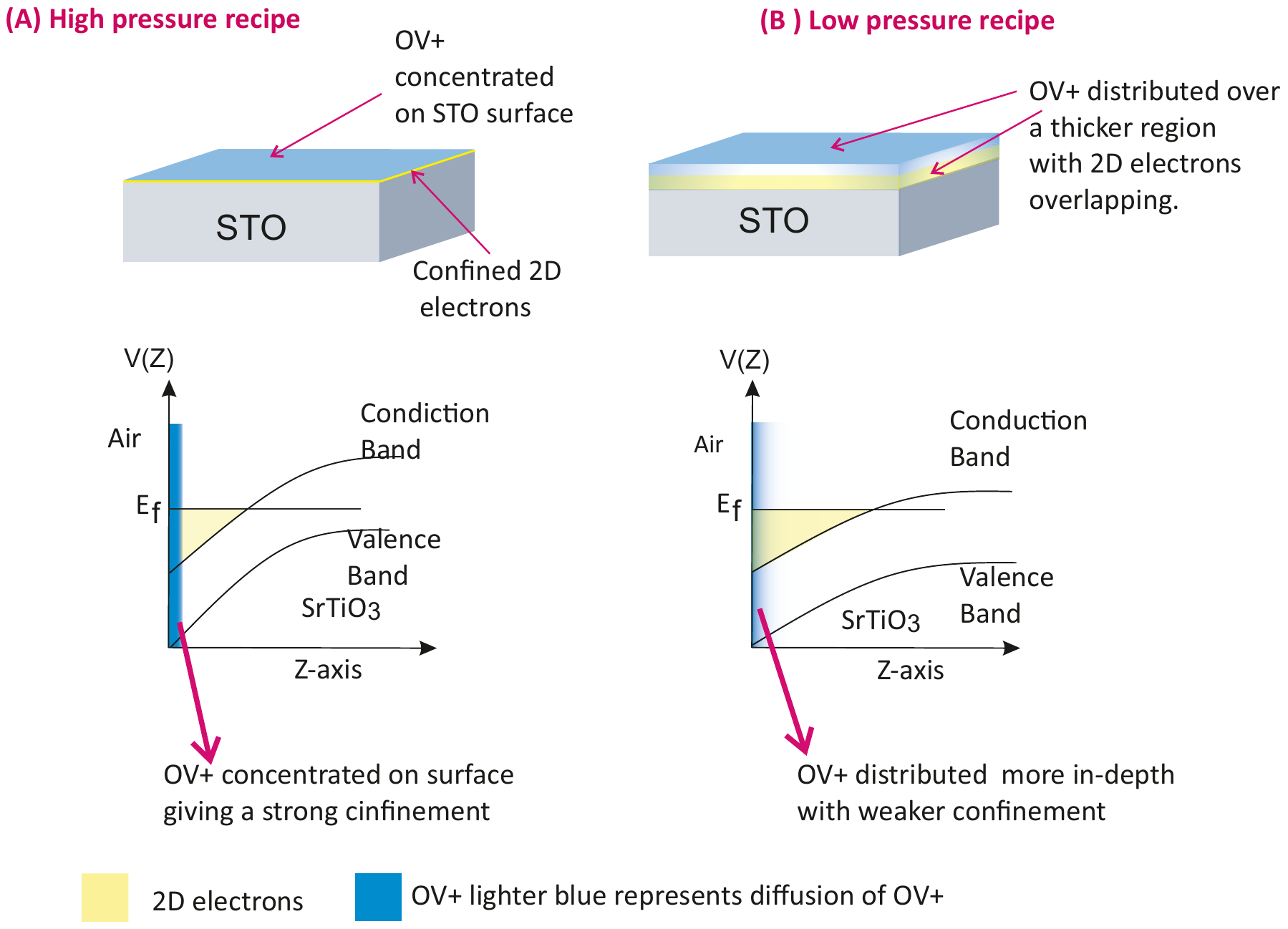}
\caption{ A simplified schematic showing the high pressure recipe which is expected to dope the surface more. The blue regions and its gradient can be viewed as diffusion of OV+ states and the yellow regions representing the $t_{2g}$ electrons forming the  q-2DES. (A) The OV+ states in  region A makes it possible to have a better confined q-2DES. The close to surface distribution removes the cubic symmetry of the crystal due to a structural inversion asymmetry that  favours an Anderson impurity type correlation amongst the OV+ states and the underlying $3d\;$-Ti orbitals  $t_{2g}$ electrons, possibly with out of plane  $ d_{xz}, d_{yz} $ orbital symmetries\cite{OV_correlated}. (B) The low pressure recipe produces a doping  profile that penetrates  deeper into the wafer with a weaker confinement. The OV+ states are in the path of the electrons in this case. In case (A) the OV+s are remote and act like dipole scatterers in modulation doping. The schematic does not show possible surface oxidation induced surface reconstructions\cite{Cathode}. }
\label{fig3b}
\end{figure}

In Argon ion bombarded samples the surfaces show an amorphous un-doped surface reconstruction of $\sim \; 3\;nm$ followed by a crystalline doped region\cite{Cathode}. We do not have any appropriate tool to study surface reconstruction in our process. But from the high mobility and transparency of the q-2DES (discussed in last section) we can state that we assume lattice distortion may not be the case in our system. It is known that OVs can be in OV+ , OV++ and a neutral state. Density functional theory (DFT) calculations have reported OV+ states that yields one electron to the system and has one bound state is most favourable\cite{OV_type2}. In doped STO the Ti$\;3d$  bands are responsible for transport. Another DFT calculation assumed a model where the itinerant electrons interact with the OV+ \cite{OV_correlated}.The OV+ state is proximate to  the Ti$\;3d_{3z^2 -r^2}$  bands and hybridized like a molecular orbital with the Fermi sea from the Ti $\;3d$ bands with $t_{2g}$  bands symmetry. For the out of plane $d_{xz}$ states are expected to have a coupling with the OV+ like an Anderson impurity. The Anderson type behavior is only possible when there are small clusters of interacting OVs with  reduced symmetry. The samples doped closer to surface at high pressures are perfect candidates for this scenario predicted by reference \cite{OV_correlated}
due to an abrupt structural inversion asymmetry  at the the oxygen deficient surface  followed by vacuum.  In Fig(\ref{fig3b}) we give a schematic of the doping profiles and confinement potentials for the high pressure and low pressure recipes.

The overall system can be non-magnetic and the local scattering of states close to the surface can give rise to a Kondo scattering.  According to reference \cite{OV_correlated} one need not have all OV states in the system acting like an Anderson impurity and a few hybridized localized states can give a clear resistive upturn with temperature from states close to surface states. The OVs are expected to be correlated via hopping of electrons from the q-2DES. Also reference \cite{OV_correlated}  pertinently points to the fact that OV+ can  can form a background image charge to stabilize a quantum well needed to trap quasi 2D surface states.

It is clear from the S-dH data that the high pressure doped samples are  closer to the surface. We cannot assume the system is dominated by impurities  as mobility is still high. Also a close to surface donor cum Anderson impurity state mimics a remote dipole like scattering in  modulation doping \cite{Stormer} without degrading mobility drastically.
The sample doped at lower pressure despite of less ionic flux has a low resistivity and higher density. We can only assume  that this is because the  energy per ion and mean free path of a low pressure plasma is higher causing OV+ to be deeper. The OV+ in isolated pockets may get easily screened especially  with high dielectric  constant re-normalization of STO.

\section*{Weak Localization and Anti-localization cross over in low pressure doped samples.}
\begin{figure}[H]
\centering
\includegraphics[width=\textwidth]{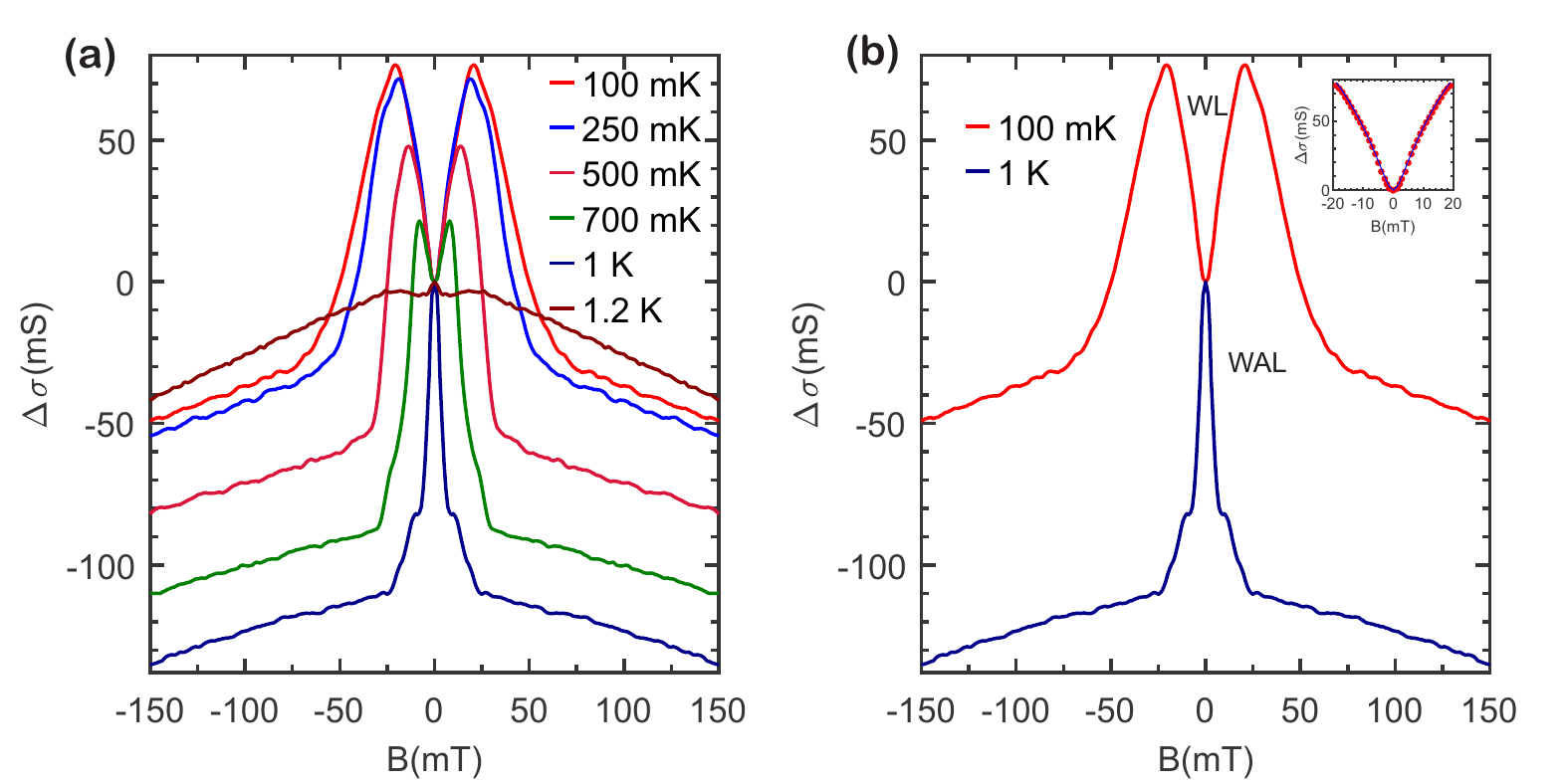}
\caption{{\small (a) Weak Localization in samples doped in low pressure. The magneto-conductivity, $\Delta\sigma=\sigma(B)-\sigma(0)$ increases with increasing temperature for a very small window of magnetic field $B$. (b) At 1K a positive  magneto-conductivity becomes negative, indicates a temperature induced crossover from weak-localization(WL) to weak anti-localization(WAL). Increasing the temperature beyond $1.2\;K$ only a negative magneto- conductance is present and indicates the destruction of quantum interference effect. In the inset a fit to the HLN formula is shown. The coherence length is independent of temperature, and is $\sim (380\pm8)\;nm$}.}
\label{fig3}
\end{figure}
Weak localization( WL) is a drop in resistance with low B fields. In a mesoscopic conductors puddles of impurities give rise to localization by coherent time-reversed interfering paths that are broken by weak magnetic fields.  Weak Anti-localization (WAL), where typically spin-orbit scattering gives rise to an increase in resistance with low magnetic fields, is a typical low field interference effect seen in STO-based systems. We do  not see either of these quantum interference effects  in the samples doped with high pressures. It is well known in systems like ultra low density Silicon q-2DES the weak localization is suppressed due to $e-e$ correlations when density is reduced\cite{WL_suppression}.  We may not have direct $e-e$ correlations in the high pressure doped systems but we can expect the source of Kondo upturn the  OV+ states to be in a correlated Anderson like coupling  with the conduction electrons\cite{OV_correlated} and de-cohere any quantum interference.
However, in the samples doped at a low pressure of 10 mTorr, we see a weak localization with a side structure at low temperatures. The same argument for Kondo applies where the local moments are  spread in a more quasi-3D region and partially screened from the conduction electrons. As discussed earlier the Hall effect also shows a weaker slope at low fields (simply implying a higher carrier density) followed by  a stronger slope at high fields. This enhancement is due to weakly paramagnetic OV+s aligning in higher fields.
A similar behavior is seen in Chromium($Cr$) doped topological insulators \cite{WAL_TI}.

As we heat the sample to $\sim\;1\;K$, we find a clear crossover from WL to WAL. Around $1.2\;K$  neither of the effects are present. As discussed before the $10\;mTorr$ doped samples are more like a broad square well in the confinement axis  with magnetic scattering impurities scattered over a larger region.  At low temperatures, the wave functions self interfere  leading to a weak localization-like behaviour flanged by a feature that looks like WAL. If we fit the standard HLN theory to WL up to the point where the Magneto-conductivity is like WL; we get the same decoherence times up to $1\;K$. Such saturation of decoherence can come from $e-e$ interactions , quantum fluctuations of two level systems of as well as magnetic impurities lowering the coherence time in metallic films and nano-structures\cite{decoherence}. Since we do not see any feature of temperature dependence and a sudden change to WAL, we conclude that WL suppresses the WAL at low temperatures. As the temperature increases, we get more scattering from the partially screened magnetic moments leading to WAL. However, both mechanisms are ineffective at higher temperatures. A similar behaviour with multiple WAL to WL to WAL crossings  is also seen in $Cr$ doped TIs \cite{WAL_TI}. Such crossovers occurring in  diverse systems  needs new theoretical models.

\section*{Optical properties and transparent optical inter connects}
In this section the optical properties  Sample-C is presented.
A metal is generally shiny and lustrous in the visible wavelength. Even in the scope of a simple Drude model, metals can have high conductivity and low reflectivity in some optical bandwidth \cite{Basov}. A correlated electron system has unique optical widows of transparent  regimes.
A strongly correlated systems like oxides displays a high wavelength for a plasma cut-off depending on the effective mass and density. A high mobility system like ours is expected to offer a functionality as a transparent conductor as effective mass $m^*$ is high. In the context of a Drude model a cut-off for a  plasma reflection wavelength can be defined as
$\lambda_p = 2\pi c \sqrt{\frac{ \epsilon_0 \epsilon_r m^* }{n e^2} }$, $\epsilon_0$ is the permittivity of free space and $\epsilon_r$ relative permittivity of the medium, and $e$ the charge on the electron. Using parameters like $m^*$ and carrier density $n$ where a thickness  $\sim 4\;nm$ was  estimated using ion penetration depth $n_{3D}$(see SI), we get a cut off as high as $\lambda_p \sim 16\;\mu m $.

\begin{figure}
\centering
\includegraphics[width=\textwidth]{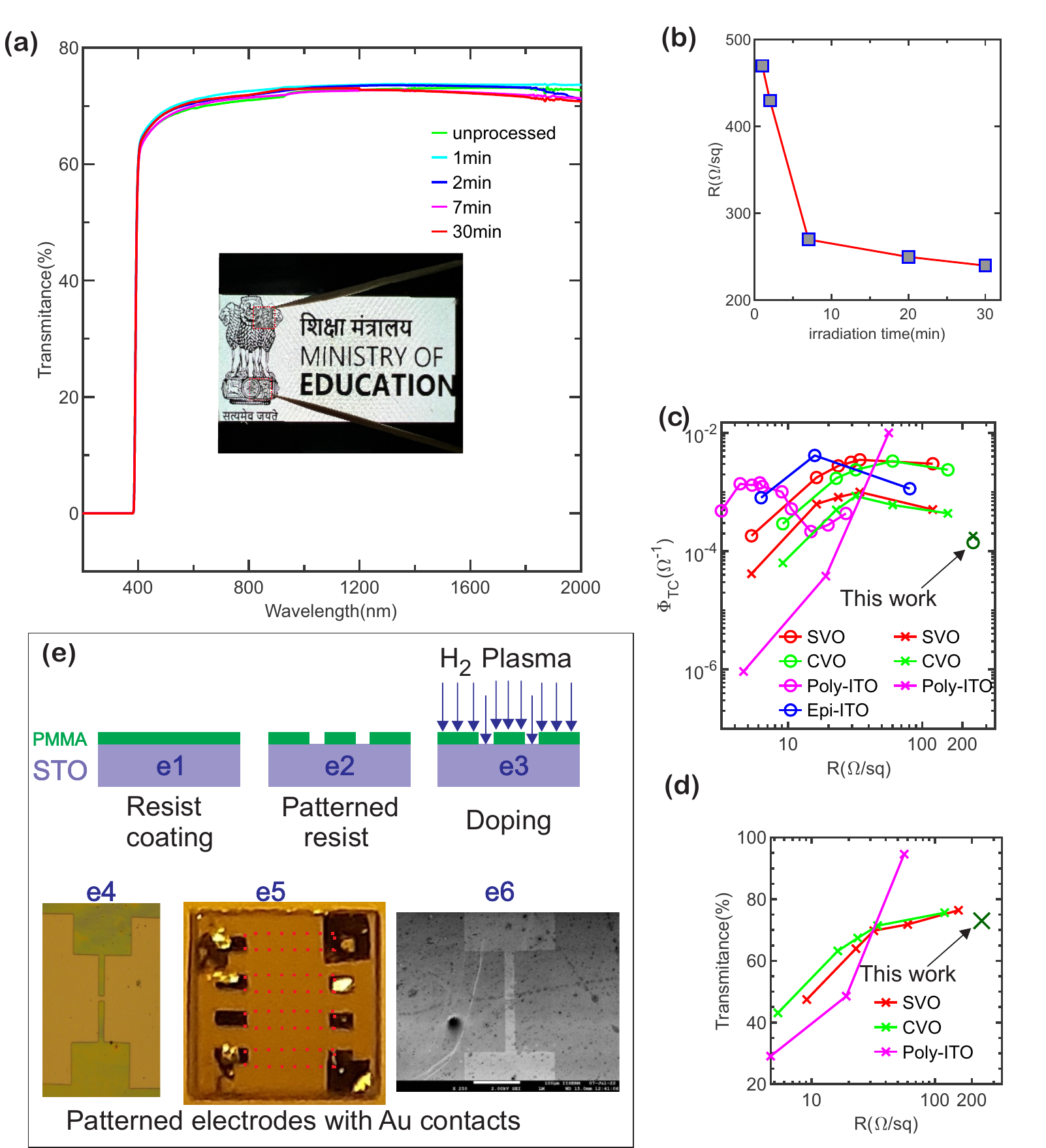}
\caption{(a) Transmission spectra for { \bf Sample-C} for different process times. The inset shows two devices {\bf Sample-C} and {\bf Sample-D } placed on a LCD display of a phone. The toothpicks point to the devices with red boundary marked on crystal edges. (b) Process time dependence of the sheet resistance. The sheet resistance is greatly reduced while transmission is still same with a non-systematic deviation  of $\pm 0.7\%$ at say for e.g., around $\lambda \sim 1.5\;\mu m $. (c) A plot of figure of merit $\Phi_{TC} = \frac{T^{10}}{R_{\square}} $ vs $R_{\square}$ for some well known set of transparent conductors from the \cite{Correlated}, data complied for $\mathrm{SrVO_3}$(SVO), $\mathrm{CaVO_3}$(CVO)(ref \cite{Correlated}), poly-ITO(ref \cite{polyITO}), epi-ITO(ref \cite{epiITO1}) versus the sheet resistance. Open circle corresponds to $550 \; nm$  and cross are for $1.5\;\mu m$ wavelength. (d) Transmission coefficient of the same set of materials and our sample around $\lambda \sim 1.5\;\mu m$, data complied for SVO, CVO(ref \cite{Correlated}), poly-ITO(ref \cite{trans1500nm}).
(e) A proof of concept device for transparent inter-connects device with standard e-beam lithography(EBL) process. (e1-e3) Process flow of lithography and doping. (e4) Optical micro-graph of processed e-beam resist. (e5) Processed device with Ti/Au contacts by lift-off.
(e6) Contrast between doped and un-doped regions in a very low magnification Scanning electron microscope image due to charging of un-doped regions. Some resist and or solvent residues were un-removable possibly as they were crossed due to electron beam exposure. }\label{fig4}
\end{figure}

Our process involving a $\mathrm{H_2}$ plasma is suitable for masking and making transparent opto-electronic devices.  We also found doping the system for several minutes down to $ R \sim \SI{240}{\ohm\per\sq}$  in a square piece of STO did not result in lowering the transparency drastically. Above UV from visible to Infrared including the telecommunication wave-length of $1.5 \;\mu m$ a negligible drop in transmission compared to a $0.5 \;mm$  thick un-doped crystal was observed.
In the first step of doping the sample was doped the sample for $60\;s$ of irradiation time. After doping optical transmission was measured in an Agilent Carey UV-VI spectrometer up-to a wavelength of $2\; \mu m$.  For the electrical measurements four Ti/Au contacts were  made at the corners of samples. Resistance was measured using DC source meter(KEITHLEY 2450 source-meter). Same sample was doped repeatedly for the different process times. All the process parameters such as bias voltage and chamber pressured were kept same for every sequential doping. As shown in figure(\ref{fig4}a), optical transparency is independent of the level of doping with a non-systematic deviation e.g,  around  $\pm0.7\%$ at $\lambda \sim 1.5 \; \mu m$. The inset of figure(\ref{fig4}a) shows sample-C and sample-D placed on a logo displayed on a LCD phone screen. The scratched edges and contacts at the peripherry are accidentally camouflaged by the pattern. Hence along with the tooth pick a red outline is included in the photograph to identify the transparent conductor.  While transparency does not change with level of doping, its conductivity is greatly enhanced and saturates above a certain level of doping. From the room temperature Hall measurement estimated sheet carrier density is $9.2\times10^{14}\;cm^{-2}$ for the highest level of doping. We found the room temperature Hall mobility was $\mu \sim21\; cm^2V^{-1}s^{-1}$ much higher than reported for Argon or LAO/STO hetero-structures.
A  figure of merit\cite{fom} for transparent conductors in terms of transmission coefficient $T$ and resistance per square is
$\phi_{TC} =\frac{T^{10}}{R_{\square}}$. With the lowest sheet resistance $\sim \SI{240}{\ohm\per\sq}$  we could achieve a value of $\phi_{TC} \sim 1.4\times 10^{-4}\;\Omega^{-1} $ at a wavelength of $550\;nm$, which is comparable to poly-crystalline Indium Tin oxide or thinner versions of other correlated materials like $\mathrm{CaVO_3}$ \cite{Correlated}. One key distinguishing feature of our samples is the figure of merit does not degrade too much  at the free space optical telecommunication wavelength of $ 1.5\; \mu m$ and above up-to our measurement limit of $ 2\;\mu m $.
These are important wavelengths for communication technologies as well as free space satellite based quantum communication protocols \cite{Qcomm}. STO is also an important substrate for superconductors like YBCO. The process can be useful in integrating thin film super-conducting nano-wire single photon detectors  (SNSPDs) or high Tc materials in SNSPds \cite{SNSPD}. \\
While our resistivity may be self limited by the doping not getting deeper but we have a clear advantage of pattern-ability of transparent conductive structures by standard lithographic techniques. We show a set of interconnects defined by standard e-beam lithography of PMMA. We show connections of different sizes as seen in the figure(\ref{fig4}e(e5)), the interconnections of different sizes ranging from 20 microns to 200 microns are  not visible.  Only the gold contacts in the end with indium bonds are visible. All connections had the same number of squares and showed a two probe resistance of  $ R\sim 12\;k\Omega $ at room temperature. In a scanning electron microscope due to charging of the un-doped regions we could see a contrast with respect to the doped regions as shown in figure (\ref{fig4}e(e6)). We could not find any significant difference in EDS as it samples deeper into the un-doped regions also and oxygen is not the heaviest element in STO. \\
These doped samples can also be enhanced in conductivity by exposure to UV. The persistent photo conductivity on exposure to few  minutes of UV stays active for several hours. This gives us one more handle to enhance the figure of merit above in the region of transparency.

To conclude we have made a novel doping process with a gentle $\mathrm{H_2}$ based plasma for $\mathrm{SrTiO_3}$. The low density doped samples clearly showed high
mobility with quantum oscillations at low temperatures. The confinement potential tune-able with pressure of the process gas leads to a Kondo like state in case of sharp
confinement potentials  and weak localization when doped with a more quasi 3D confinement. Our process also yields a new type of transparent conductor that can be used to pattern interconnects and various opto-electronic devices. The pattern-ability amenable to standard lithographic processes is a
key feature that stands out from MBE or PLD techniques alone for device applications up-to telecommunication wavelength and beyond. Un-doped $\mathrm{SrTiO_3}$ or similar $\mathrm{ABO_3}$ layers can be grown by MBE or PLD on other substrates. Unlike blanket doping of  films during growth  our technique can selectively dope regions by lithography.  When combined with different substrates and anti-reflective coatings the possibilities of novel opto-electronic devices are manifold.
The high mobility we have achieved also makes it a new system for investigating nano-fabricated devices in a system where electronic properties both spin and charge  can be combined with novel dielectric properties of STO \cite{ferrospin} can be useful in new devices.
In conclusion we are able to Engineer OV+s to produce a strongly confined as well as  correlated hybridized Anderson impurity type Kondo system or weakly confined conductor where correlations of OV+ are screened.
The tightly confined system with correlations  naturally yields pattern-able transparent electrodes for opto-electronic devices above UV into the telecom wavelength.

\section*{Conflict of Interest}
The Authors declare no conflict of interest.
\section*{Data Availability}
The raw data and processed data will be made available with the pre-print.
\section*{Acknowledgements}
We thank IISER Mohali established under Ministry of Education for support.
We thank DST-nano-mission project no. SR/NM/NS-1098/2011. We thank DST-FiST for supporting studies in this direction with equipment like Wire bonder and  PCB fabrication system.  We thank the SEM facility at IISER Mohali for SEM images and lithography. We thank Department of Chemical Sciences Instrumentation facility at IISER Mohali for access to UV-VIS spectrometers. We thank Prof B. Muralidhran IIT Bombay, and Dr. Debangsu Roy IIT Ropar for reviewing the manuscript critically. We thank Dr. Subrata Ghosh IIT Mandi for advising us on appropriate patent procedures before a pre-print. PK thanks CSIR  for funding.

\section*{Author Contribution}
SSY, SK, PK did the measurements under guidance of AV. SSY did the analysis with inputs and help from other authors. AV wrote the manuscript with inputs from all the authors.
\bibliographystyle{unsrtnat}
\bibliography{shubnikov_biblio}
\end{document}


\date{}
	\maketitle
	\section{Sample fabrication}
	
	STO wafers oriented on $(100)$ planes were procured  pre-cut or scribed  into a $5mm\times5mm$  size with $\mathrm{TiO_2}$ termination from Shinkosha Corp. To pattern a Hall bar geometry for sample A, we have designed a stencil mask made out of stainless steel with a minimum  feature size of $\sim 150\; \mu m$ by chemical etching used for solder masks. 
	\begin{figure}[ht]
		\centering
		\includegraphics[width=\textwidth]{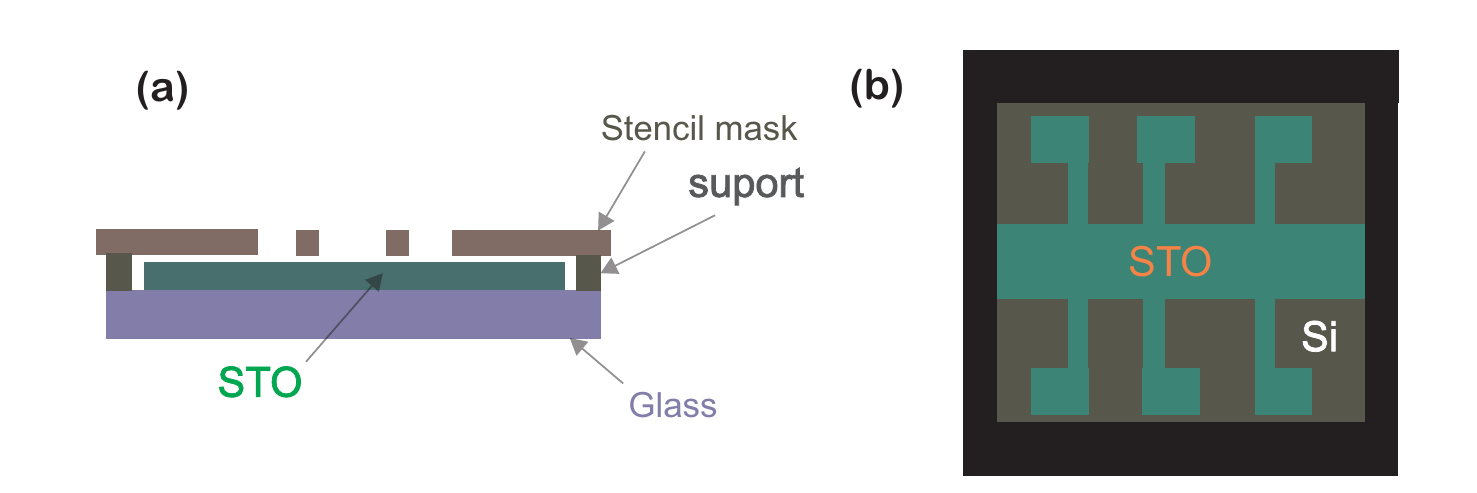}
		\caption{Device fabrication(a) A stencil mask holding on a glass for Hall bar fabrication. (b) Schematic of the Hall bar with edges painted with nail varnish.}
		\label{figureS1}
	\end{figure}
	The Hall bar had the $R_{xx}$ voltage probes enclosed between $1400 \mu m $ and $600 \mu m$ wide. The STO(100) wafer was mounted on a glass substrate using PMMA; two silicon wafers were put close to the STO wafer to hold the stencil mask and also to avoid any undesired scratch on the STO. A $45 - 50\;nm$ un-doped silicon was evaporated using an electron beam evaporator with substrate held at ambient temperature. A test sample where there was a silicon line  deposited in between the hall bar showed overload across while two contacts on the same sides worked. Hence the silicon stencil did not reduce the STO.  In figure(\ref{figureS1}) schematic of the fabrication process is shown. The key step is to have a thicker silicon wafer to hold the mask without scratching the STO. 
	
	Edges of the sample were painted using insulating nail lacquer and dried overnight. Samples were loaded to the Reactive Ion Etching(RIE) system, and the chamber was pumped for  atleast 2 hours before irradiation. The RIE system was equipped with a mass flow controller with a maximum of $99 \; \textrm{sccm}$ for hydrogen, and chamber pressure can be controlled up to a maximum of $\sim 95 \;mTorr$. A 13.5 MHz RF source with an automated matching network is used to create a dense plasma. After irradiation, Au/Ti contacts were evaporated thermally for the electrical contacts through a second stencil mask window. For sample B whole surface was doped except for the  edges protected by nail varnish. $\mathrm{Au/Ti}$  contacts were deposited using a stencil mask. Scratches were made manually to avoid the contacts drawing currents as shown in inset of fig1 in main text.
	\newpage
	\section{Temperature dependence of Transport  Data} 
	The temperature dependence of the symmetrized $R_{xx}$ and the anti-symmetrized $R_{xy}$ from base temperature to $T\sim 1\; K $ is shown here. 
	\begin{figure}[h]
		\centering
		\includegraphics{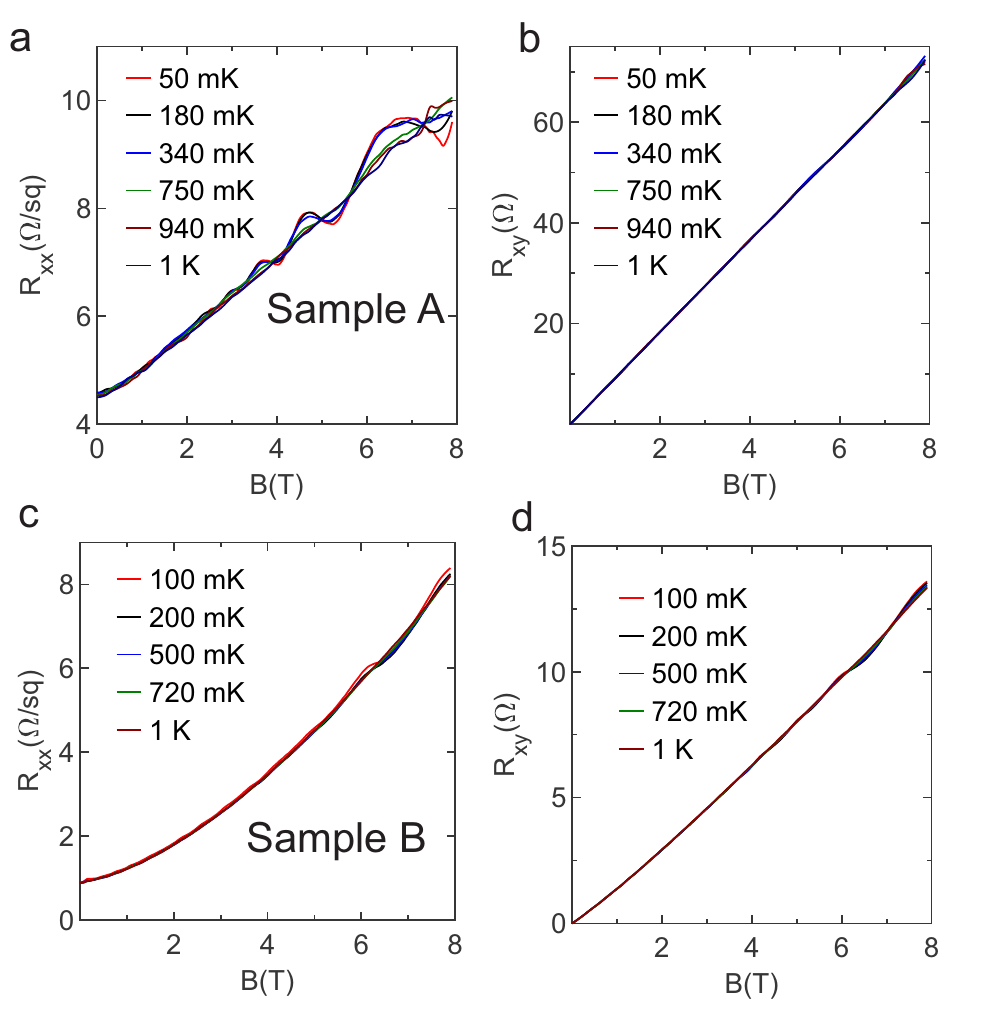}
		\caption{Temperature dependence of the magneto resistance. (a) Symmetric component, $R_{xx}$  and (b) anti symmetric component, $R_{xy}$ for sample A. (c) and (d) are same for sample B.}
		\label{figureS2}
	\end{figure}
	
	\section{Data from additional sample}
	\begin{figure}[ht]
		\centering
		\includegraphics[scale=.7]{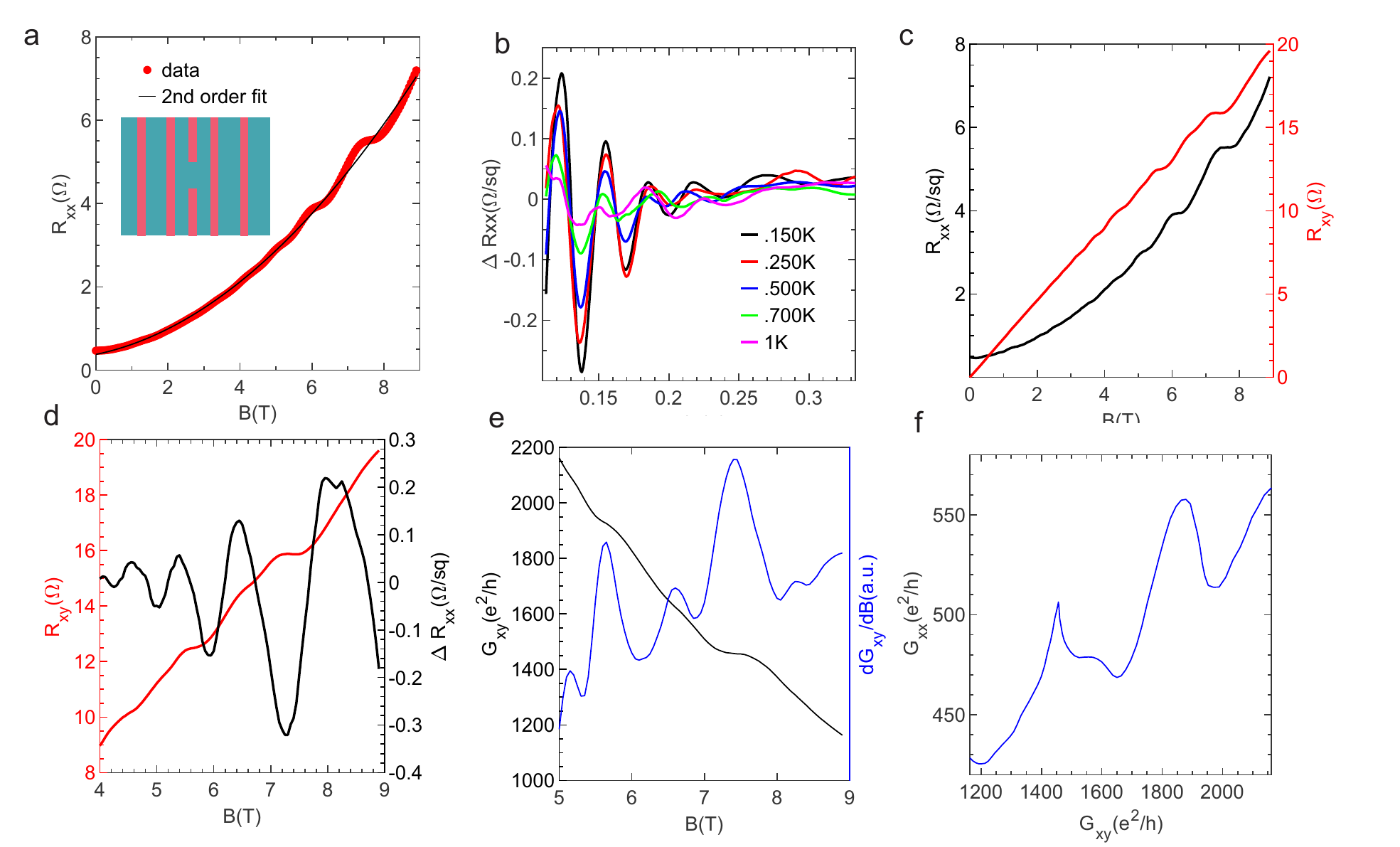}
		\caption{Transport measurement in another sample. (a) Longitudinal resistance $R_{xx}$ vs magnetic field measured at $150\;mK$. Black soild line is a fit using polynomial, $a_0+a_1B+a_2B^2$, $a_0,\;a_1$ and $a_3$ are the polynomial coefficients. In the inset measurement configuration is shown. (b) SdH oscillations at different temperatures. (c) A weak plateau like structure in Hall resistance. (d) Hall plateaus plotted with oscillating longitudinal resistance. (e) Hall conductance $G_{xy}$ and derivative. Position of the Hall plateaus can occure at the local maxima in derivative, as mention in ref\cite{XIE201425}, and (f) a plot of $G_{xy}=\frac{R_{xy}}{(R_{xx}^2+R_{xy}^2)}$ and longitudinal conductance $G_{xx}=\frac{R_{xx}}{(R_{xx}^2+R_{xy}^2)}$.}
		\label{figureS3}
	\end{figure}
	In the main, text we have presented the data for samples A and B. In figure(\ref{figureS2}), we have shown the data for the sample, which was measured before measuring A and B. For this whole wafer was irradiated in $\mathrm{H_2}$ plasma for $30\;sec$ at a bias voltage of $100\;V$, then aluminium contacts were made using a stencil mask, as shown in the inset of the figure(\ref{figureS3}a). Estimated low-temperature mobility was $23700\;cm^{2}V^{-1}s^{-1}$ and density $2.6\times10^{14}\;cm^{-2}$. The pressure  was lower by 5 mbar from the set value and reached 70 mTorr during the process. Hence it is not reported as data in the main text. The other samples were stabilized to the desired pressure value for 5 minutes  and then the plasma was ignited. A second-order polynomial is used to remove the classical contribution of the magnetoresistance; the oscillating part of the resistance is shown in figure(\ref{figureS2}) against the inverse of magnetic field $\frac{1}{B}$. A weak quantum Hall-like plateau at very high filling factors is seen in the $R_{xy}$ as reported in reference \cite{XIE201425}. 
	\section{HLN fitting}
	A quantitative estimation of the phase coherence length  can be found using the Hikami-Larkin-Nagaoka(HLN) formula, and is given as\cite{hln},
	\begin{equation}
		\Delta\sigma(B)=\alpha\frac{e^2}{2\pi^2\hbar}{\left [ ln(\frac{B_\phi}{B}) -\psi(\frac{1}{2}+\frac{B_\phi}{B})\right ]},
		\label{hln}
	\end{equation}\\
	where $\Delta\sigma(B)=\sigma(B)-\sigma(0)$, is the magneto-conductivity, $B_\phi$ is a characteristic de-phasing field, and $\psi$ is a Digamma  function. Here $\alpha$ is a interaction dependent parameter. $B_\phi$ is a characteristic dephasing  magnetic field and is related to phase coherence length $l_\phi$  as, $B_\phi=\hbar/4el_\phi^2$. Using Eq(\ref{hln}) a $l_\phi$ of the $(380\pm20)\;nm$ is found, and is independent of the temperature. 
	\begin{figure}[h]
		\centering
		\includegraphics[scale=1]{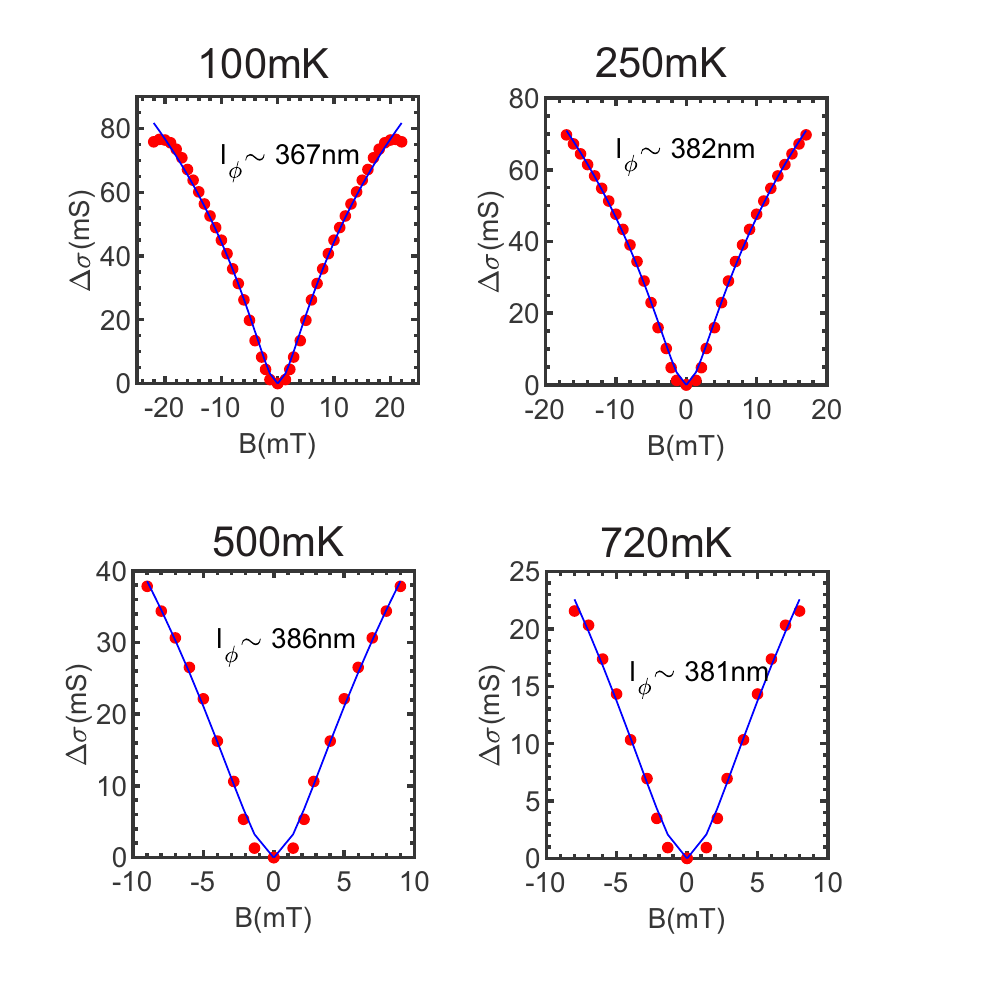}
		\caption{HLN fitting for different temperature. Data were fitted using eq(\ref{hln}). Estimated average phase coherence length is, $l_\phi\simeq380\;nm$.}
		\label{figureS4}
	\end{figure}
	\section{Kondo model}
	To get the deep insight we further analyzed the data using Kondo model. Sheet resistance due to  all possible mechanism is given by Eq.~(\ref{kondo model}) 
	\begin{equation}
		R^{fit}\left ( T \right )=R_{0}+qT^{2}+pT^{5}+R_{K}T\left ( \frac{T}{T_{K}} \right ) \label{kondo model}.
	\end{equation}
	Here $R_{0}$ is the total residual resistance due to sample disorder and are temperature independent, and $T^{2}$ and $T^{5}$ terms are temperature dependent contributions from  the electron-electron and electron-phonon interactions respectively. The last term comes from the Kondo scattering. For the model fitting to the data we have used the empirical form of the universal resistivity function \cite{Lee,Goldhaber}
	\begin{equation}
		R_{K}\left ( \frac{T}{T_{K}} \right )=R_{K}\left ( T=0 \right )\left ( \frac{T_{K}^{'2}}{T^{2}+T_{K}^{'2}} \right )^{s},
	\end{equation}
	where, $R_K(T=0)$ is the Kondo resistance at zero temperature, and $T_{K}^{'}=T_{K}/\left ( 2^{1/s}-1 \right )^s$. For the numerical fitting, $s$=.225 which was obtained from the re-normalization group is used\cite{Goldhaber}. We used this model in the main text to fit the observed Kondo like behaviour.
	
	\section{Optical properties} 
\begin{figure}[h]
		\centering
		\includegraphics[scale=.9]{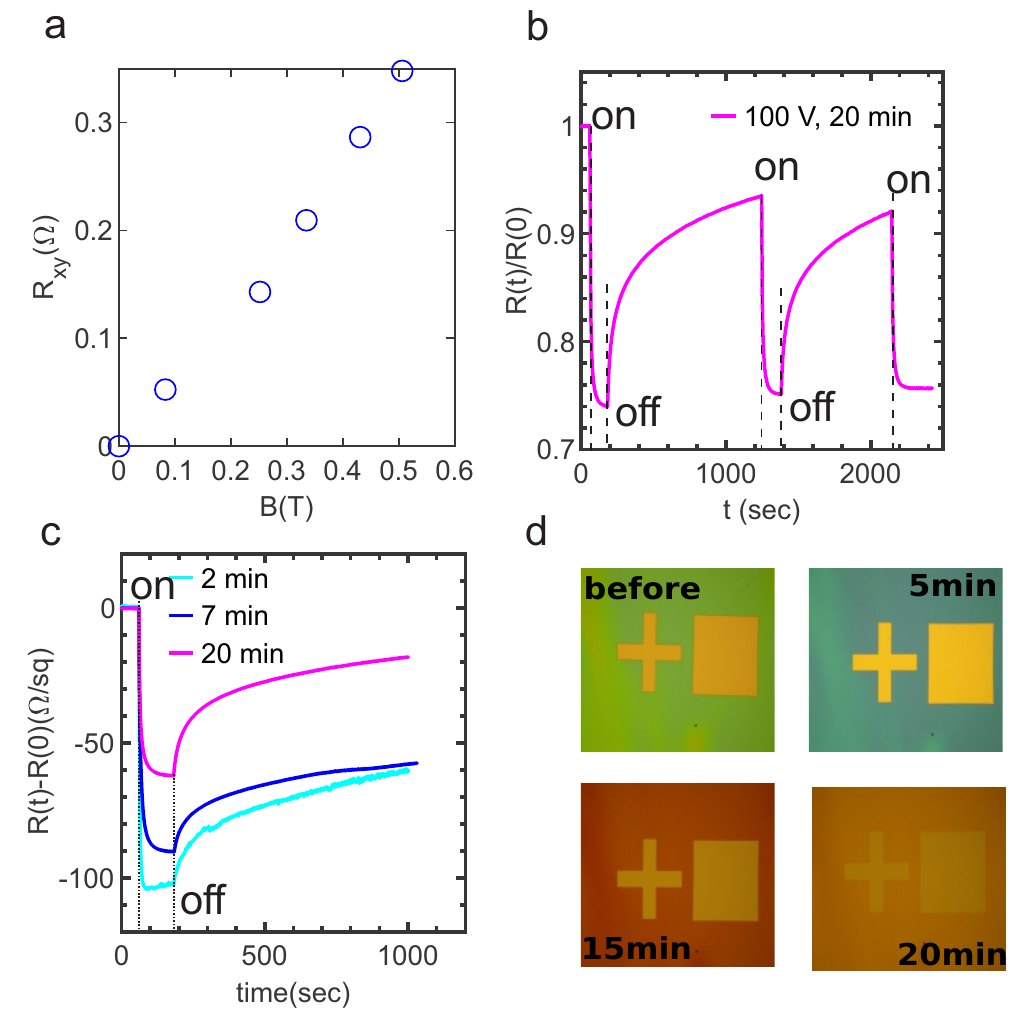}
		\caption{Optical properties of the heavily doped STO. (a) Hall effect at room temperature. (b) Resistance switching due to on and off of a UV light(wavelength=$367\;nm$). (c) Persistence of photoconductivity for different irradiation time, and (d) Resist(PMMA) survival in a $\mathrm{H_2}$ plasma. Patterned structures can survive($\sim100\;nm$ thick resist) for a very log irradiation time.}
		\label{figureS5}
	\end{figure}

In figure(\ref{figureS5}), we have shown room temperature properties of the heavily doped sample. For the hall measurements, we have used the Van der Pauw  method, where four contacts were made at the corners of the doped surface. A resistance switching can be seen under shining the UV light(wavelength$=367\;nm$) as shown in figure(\ref{figureS5}b). This can happen due to the enhanced oxygen vacancies and excitation of the trapped electronic states. The photo-conductivity persists for a long time. As can be seen in figure(\ref{figureS5}c), photo-conductivity does not decrease abruptly but persist. We pattern cross marks using e-beam lithography to test the resist survival rate. A PMMA with a thickness $\sim\;100\;nm$ can survive for a long period of irradiation, allowing direct patterning of the doped device.  
	A device fabrication process was tested by patterning the electrodes with different widths using a standard electron beam lithography technique. A $100\;nm$ thick PMMA resist was coated on a clean un-doped STO. Exposed patterns  were developed by a developer. Then the sample was doped using $H_2$ plasma irradiation. An optical image was taken just after doping, and Au/Ti contacts were deposited by covering the main structure with an aluminium foil. Then the resist was removed in an Acetone solvent.
	\newpage
	\section{Thickness estimation of the doped surface}	
	We can estimate the rough thickness of the q-2DES by calculating the penetration depth of the ions within STO.  As we already mentioned, oxygen vacancies are the primary mechanism to doped the surface, and the low energy interaction of ions induces these vacancies to the surface of the STO.We can say the ion's depth could be the doped region's rough thickness. The empirical form for the ion penetration length in an ion matter interaction is given as\cite{Reagor2005,Harper}
	\begin{equation}
		L=1.1\frac{E^{2/3}W}{\rho(Z_{i}^{1/4}+Z_{t}^{1/4})^2}
	\end{equation}
	where $E$ is the energy in eV, $W$ is the atomic weight of the target in atomic mass units, $\rho$ is the target density, $Z_{i,t}$ are the atomic numbers of the ion and target, respectively. Here the target is a compound so we take the weighted average of the atomic weights and numbers to calculate the penetration depth. The estimated thickness is found to $\sim3.6\;nm$. Here we have not included the effect of pressure and irradiation time in the formula, in general, low pressure will dope deeper due to the high mean free path of the ions\cite{Harper}.
	\bibliographystyle{unsrtnat}
	\bibliography{sup_bib}